\documentclass[pra,aps,groupaddress,showpacs,twocolumn]{revtex4}
\usepackage{epsfig,amsmath,graphicx,amssymb}
\def\be{\begin{equation}}
\def\ee{\end{equation}}
\def\bea{\begin{eqnarray}}
\def\eea{\end{eqnarray}}
\def\bes{\begin{subequations}}
\def\ees{\end{subequations}}
\begin{document}
\title{Highly Entangled Photons and Rapidly Responding Polarization Qubit Phase
Gates in a Room-Temperature Active Raman Gain Medium}

\author{Chao Hang$^{1,2}$ and Guoxiang Huang$^{1,3 \footnote{Electronic address: gxhuang@phy.ecnu.edu.cn}} $}
\affiliation{$^1$State Key Laboratory of Precision Spectroscopy and
Department of Physics, East China Normal University, Shanghai
200062, China \\
$^2$Centro de F\'isica Te\'orica e Computacional, Universidade de
Lisbon, Complex Interdisciplinary, Avenida Professor Gama Pinto 2,
Lisbon 1649-003, Portugal \\
$^3$Institute of Nonlinear Physics, Zhejiang Normal University,
Jinhua, 321004 Zhejiang, China}

\date{\today}

\begin{abstract}

We present a scheme for obtaining  entangled photons and quantum
phase gates in a room-temperature four-state tripod-type atomic
system with two-mode active Raman gain (ARG). We analyze the
linear and nonlinear optical response of this ARG system and show
that the scheme is fundamentally different from those based on
electromagnetically induced transparency and hence can avoid
significant probe-field absorption as well as temperature-related
Doppler effect. We demonstrate that highly entangled photon pairs
can be produced and rapidly responding polarization qubit phase
gates can be constructed based on the unique features of enhanced
cross-phase modulation and superluminal probe-field propagation of
the system.

\end{abstract}


\maketitle

\section{Introduction}

Efficient schemes for producing entangled photons and constructing
all-optical quantum gates are very important in optical quantum
information processing and computation \cite{Nielsen}. For this
aim, a significant suppression of optical absorption and a giant
enhancement of Kerr nonlinearity is crucial. However, in a
conventional medium this can not be efficiently implemented
because optical fields far away from atomic resonance are used to
avoid large optical absorption, and hence the Kerr nonlinearity of
the system is usually very weak.

In recent years, much attention has been paid to the study of
electromagnetically induced transparency (EIT) in resonant atomic
systems \cite{Fleischhauer,ham}. The wave propagation in EIT media
possesses many striking features, such as the large suppression of
optical absorption, the significant reduction of group velocity,
and the giant enhancement of Kerr nonlinearity
\cite{Fleischhauer}. Based on these features, many EIT-based
applications, including optical quantum memory \cite{lvo},
high-efficient multi-wave mixing \cite{Fleischhauer}, optical
atomic clocks \cite{san,zan,hong}, and slow-light solitons, etc.
\cite{Wu,huang,Michinel,hang}, have been intensively studied.
Moreover, EIT-based schemes for producing entangled photons
\cite{Luk,Peng,fri} and polarization qubit quantum phase gates
(QPGs) \cite{Ottaviani,reb,jos,Hang} have also been proposed.
However, the EIT-based schemes have some inherent drawbacks, such
as the probe attenuation and spreading at room temperature and the
long response time due to the character of ultraslow propagation
\cite{note}. These drawbacks impede the potential applications of
EIT media for rapidly responding all-optical devices at
room-temperature.

In this work, we shall propose a scheme to realize highly efficient
entangled photons and rapidly responding polarization QPGs in a
resonant atomic system. The new scheme is based on active Raman gain
(ARG) (or gain-assisted) configurations, which was demonstrated to
be able to produce stable superluminal propagations of optical waves
\cite{Chiao,stein,wang,sten,aga,jia}. Contrary to the EIT-based
schemes where the probe field operates in an absorption mode, the
key idea of the ARG-based schemes is that the probe field operates
in a stimulated Raman emission mode. Thus, they can avoid to be
affected by a temperature related Doppler effect and significant
probe field attenuation or distortion. Recently, it has been shown
by Deng {\it et al.} \cite{Deng,jia1} that large and rapidly
responding cross-Kerr effect are possible in ARG-based media. In
addition, superluminal optical solitons are also predicted in such
systems \cite{Huang,hang1}. Our system suggested here is a
four-state tripod-type atomic one with a two-mode pump field and two
weak fields. We shall prove that the unique features of the present
system can be used to produce highly entangled photon pairs and
implement rapidly responding polarization QPGs. Contrary to the
entangled photons and QPGs in EIT media
\cite{Ottaviani,reb,jos,Hang}, the present ARG scheme has the
following advantages: (i) It is able to eliminate the significant
probe attenuation and distortion induced by temperature related
Doppler effect, hence we can produce entangled photons with high
degree and implement QPGs with high reliability at room temperature;
(ii) It allows superluminal wave propagation, hence one can
implement QPGs with very rapid response. The results presented in
this work may be useful for guiding related experiments and
facilitating practical applications in quantum information science
\cite{Boyd}.

The paper is organized as follows. In the next section we give a
description of the model under study, and present the expressions
of electric susceptibilities and the group velocity of probe and
signal fields. In Sec. III, we describe a method to produce
entangled superluminal photons and construct polarization QPGs
based on the present ARG system. In the last section, we provide a
simple discussion on the temperature related Doppler effect and
quantum noises. The main results of our research are also
summarized.

\section{The Model and linear and nonlinear susceptibilities}

We start with considering a life-time broadened four-level
tripod-type atomic gas interacting with a strong continuous-wave
two-mode pump laser field (with electric fields $E_{P1}$ and
$E_{P2}$), and two weak, pulsed laser (probe and signal) fields
(with electric fields $E_{p}$ and $E_{s}$), as shown in Fig. 1.
The pump fields $E_{P1}$ and $E_{P2}$ are of $\pi$-polarization
and couple the ground state $|1\rangle$ to the excited state
$|2\rangle$ with large one-photon detunings $\delta_1$ and
$\delta_1+\Delta$ ($|\Delta|\ll|\delta_1|$), respectively. The
probe (signal) field $E_p$ ($E_s$) is of $\sigma^+$
($\sigma^-$)-polarization and couples the excited state
$|2\rangle$ to the hyperfine state $|3\rangle$ ($|4\rangle$) with
a two-photon detuning $\delta_p$ ($\delta_s$). The system contains
two Raman resonances due to the two-mode pump field for each weak
field.
%
\begin{figure}
\centering
\includegraphics[scale=0.4]{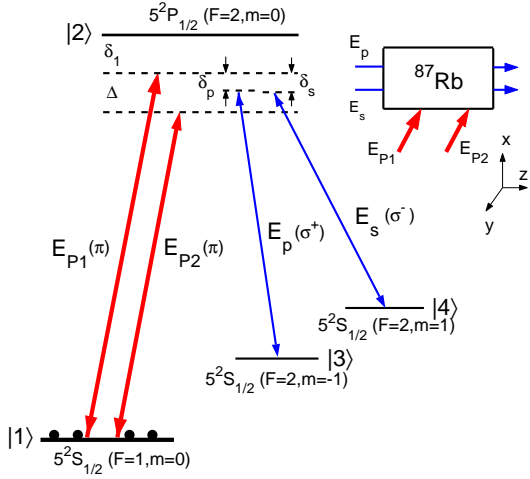}
\caption{\footnotesize (Color online) The energy levels
$|l\rangle$ ($l$=1-4) and excitation scheme of the life-time
broadened four-state tripod-type atomic system interacting with a
strong continuous-wave two-mode pump laser field (with electric
fields $E_{P1}$ and $E_{P2}$) and two weak, pulsed (probe and
signal) fields (with electric fields $E_{p}$ and $E_{s}$).
$E_{P1}$ and $E_{P2}$ are of $\pi$-polarization, while $E_p$
($E_s$) is of $\sigma^+$- ($\sigma^-$-) polarization. $\delta_1$,
$\delta_p$, $\delta_s$, and $\Delta$ are detunings. The inset
shows a possible geometry of experimental set.}
\end{figure}
Our scheme can be realized by a specific implementation using the
D1 line of $^{87}$Rb, where a homogeneous magnetic field parallel
to the laser propagation direction is applied to encode binary
information and avoid the undesirable couplings. A possible
geometry of experimental arrangement is suggested in the inset of
the figure. Note that the system we are considering here is a
direct extension (by adding a new, weak signal field) of that used
by Wang {\it et al}. \cite{wang} for the remarkable observation of
stable, superlumninal light propagation in an ARG system.

The evolution equations for the atomic probability amplitudes
$a_l(t)$ ($l$=1-4) are
\bes \label{AV}
\bea
\dot{a}_1&=& \frac{\gamma_1}{2}a_1+i\Omega_{P1}^{\ast}e^{i\delta_1
t}a_2+i\Omega_{P2}^{\ast}e^{i(\delta_1+\Delta)
t}a_2, \\
\dot{a}_2&=& -\frac{\gamma_2}{2}a_2+i\Omega_{P1}e^{-i\delta_1
t}a_1+i\Omega_{P2}e^{-i(\delta_1+\Delta)
t}a_1 \nonumber\\
& & +i\Omega_{p}e^{-i(\delta_1+\delta_p)
t}a_3+i\Omega_{s}e^{-i(\delta_1+\delta_s) t}a_4, \\
\dot{a}_3&=& -\frac{\gamma_3}{2}a_3
+i\Omega_{p}^{\ast}e^{i(\delta_1+\delta_p)
t}a_2, \\
\dot{a}_4&=& -\frac{\gamma_4}{2}a_4
+i\Omega_{s}^{\ast}e^{i(\delta_1+\delta_s) t}a_2,
\eea
\ees
where $\Omega_{Pn}=-D_{21}{\cal E}_{Pn}/(2\hbar)$ ($n=1$, 2),
$\Omega_{p}=-D_{23}{\cal E}_{p}/(2\hbar)$, and
$\Omega_{s}=-D_{24}{\cal E}_{s}/(2\hbar)$ are half-Rabi
frequencies for $|1\rangle\leftrightarrow|2\rangle$,
$|3\rangle\leftrightarrow|2\rangle$, and
$|4\rangle\leftrightarrow|2\rangle$ transitions, with relevant
electric dipole moments $D_{21}$, $D_{23}$, and $D_{24}$, and
electric-field envelopes ${\cal E}_{Pn}$,  ${\cal E}_{p}$, and
${\cal E}_{s}$, respectively. The detunings are defined by
$\delta_1=\omega_{21}-\omega_{P1}$,
$\Delta=\omega_{21}-\omega_{P2}-\delta_1$,
$\delta_p=\omega_{23}-\omega_{p}-\delta_1$, and
$\delta_s=\omega_{23}-\omega_{s}-\delta_1$ (see Fig. 1).
$\gamma_1$ presents the gain of state $|1\rangle$ for describing
the effect of atoms going back to the ground state before being
exited again. $\gamma_l$ ($l=2-4$) present the decay rates of
state $|l\rangle$ for describing the effects of both spontaneous
emission and dephasing. In the present work, we are interested in
a closed system, i.e., there is no decay to levels outside the
system we study, and hence $\gamma_1$ can be determined by the
decay rates of higher states $\gamma_l$ ($l=2-4$) through the
conservation of particle number $\sum_{l=1}^4|a_l|^2=1$ [see Eq.
(\ref{Gama}) below]. Notice that here we employ the amplitude
variable approach for the description of the motion of atoms and
$\gamma_l$ are introduced in a phenomenologically manner. A
complete description to include spontaneous emission and dephasing
can be obtained by a density-matrix equation approach. However,
for the ARG-based coherent atomic systems, two approaches are
equivalent.

In order to investigate the propagation of the probe and signal
fields, Eqs. (\ref{AV}) must be solved simultaneously with the
Maxwell equation. With the electric-field defined by $E_{j}={\cal
E}_{j}\exp [i(k_j-\omega_j t)]$+c.c., we obtain
\be\label{MAX} i\left(\frac{\partial }{\partial
z}+\frac{1}{v_g^{j}}\frac{\partial }{\partial t}\right){\cal
E}_{j}+\frac{\omega_{j}}{2c}\chi_{j}{\cal E}_{j}=0,\quad (j=p,\,s)
\ee
obtained under slowly-varying amplitude approximation, where
$v_g^{j}$ is the group velocity, generally defined as
$v_g^{j}=c/(1+n_g^{j})$ with $n_g^{j}=\text{Re}[\chi_j]/2
+(\omega_j/2)(\partial\text{Re}[\chi_j]/\partial\omega)|_{\omega=\omega_j}$
being the group index. Susceptibilities of the two weak fields are
defined by $\chi_{p,s}={\cal N}_aD_{0}a_2a_{3,4}^{\ast}/(\epsilon_0
{\cal E}_{p,s})$ ($D_{23}\simeq D_{24}=D_{0}$), with ${\cal N}_a$
the atomic concentration.

We assume that atoms are initially populated in the ground state
$|1\rangle$. For large one-photon detunings $\delta_1$ and
$\delta_1+\Delta$ the ground-state depletion is not significant,
i.e. $a_1\simeq1$. However, in order to take into account the
nonlinear effect, we need to consider the higher-order contribution
of $a_1$, which can be obtained by using the condition
$\sum_{i=1}^{4}|a _{i}|^2=1$. Meanwhile, we assume that the typical
temporal duration of the probe and signal fields is long enough so
that we can solve the equations adiabatically. With these
considerations, we obtain the expressions of $\gamma_{1}$ and
electric susceptibilities of the system
\bea\label{Gama}
\gamma_1&=&\gamma_2(G_1+G_2)+\gamma\left[\frac{G_1}{\delta_2^2}
+\frac{G_2}{(\delta_2-\Delta)^2}\right]\nonumber\\
& & \times(|\Omega_{p}|^2+|\Omega_{s}|^2),
\eea
and
\be\label{Chi1}
\chi(\omega_{j})\simeq\chi_{j}^{(1)}+\chi_{j}^{(3,s)}| {\cal
E}_{j}|^2+\chi_{j}^{(3,c)}|{\cal E}_{j'}|^2,
\ee
with $j,\,j'=p,\,s$ ($j\neq j'$) and
\bes \label{Chi2}
\bea
& & \label{Chi21} \chi_{j}^{(1)} \simeq
-\kappa\left(\frac{G_1}{\delta_2-i\gamma/2}
+\frac{G_2}{\delta_2-\Delta-i\gamma/2}\right),\\
& & \label{Chi22} \chi_{j}^{(3,s)}
=\chi_{j}^{(3,c)}\simeq\kappa'\left(\frac{G_1}{\delta_2
-i\gamma/2}+\frac{G_2}{\delta_2-\Delta-i\gamma/2}\right)\nonumber\\
& & \hspace{2.5cm}\times
\left[\frac{G_1}{\delta_2^2}+\frac{G_2}{(\delta_2-\Delta)^2}\right].\quad
\eea
\ees
Here, $\chi_{j}^{(1)}$, $\chi_{j}^{(3,s)}$, and $\chi_{j}^{(3,c)}$
determine the linear, self-, and cross-Kerr nonlinear responses of
the system. The constants in (\ref{Gama}) and (\ref{Chi2}) are
defined by $G_1=|\Omega_{P1}|^2/\delta_1^2$,
$G_2=|\Omega_{P2}|^2/(\delta_1+\Delta)^2$, $\kappa={\cal
N}_a|D_{2}|^2/(\hbar\epsilon_0)$, and $\kappa'={\cal
N}_a|D_{2}|^4/(\hbar^3\epsilon_0)$. We should also mention that in
order to obtain simplified expressions of $\gamma_1$ [i.e. Eq.
(\ref{Gama})\,] and third-order susceptibility [i.e. Eq.
(\ref{Chi22})\,], we have taken $\delta_p=\delta_s=\delta_2$ and
$\gamma_3\simeq \gamma_4=\gamma$, and used the conditions
$\gamma_2^2\ll\delta_1^2$, $\gamma^2\ll\delta_2^2$,
$\gamma^2\ll(\delta_2-\Delta)^2$, and $G_{1,2}\ll1$.  The real and
imaginary parts of $\chi_{j}^{(1)}$ denote the phase shift per unit
length and absorption or gain, respectively. From the expression of
Eq. (\ref{Chi22}), we see that the linear susceptibility for both
the probe and signal fields have two Raman resonances, which
contribute from two pump fields. If $\delta_2=\Delta/2$ and the
intensities of the two pump fields are well adjusted so that
$G_1=G_2=G$, one has $\rm{Re}[\chi_{j}^{(1)}]=0$, and hence a
gain-dependent linear phase can be completely removed \cite{Deng}.
In this case, $2\rm{Im}[\chi_{j}^{(1)}]\simeq-8\kappa
G\gamma/\Delta^2$ describes the intensity gain acquired by two weak
fields. This is fundamentally different from all EIT-based systems
which are inherently absorptive. The above choice of two-mode pump
intensities and two-photon detuning also yields
$\rm{Re}[\chi_{j}^{(3)}]=0$ and
$2\rm{Im}[\chi_{j}^{(3)}]\simeq64\kappa' G^2\gamma/\Delta^4$, i.e. a
zero nonlinear phase shift and a nonzero nonlinear intensity
absorption. Therefore, in order to obtain a nonzero nonlinear phase
shift, we need to slightly disturb the conditions
$\delta_2=\Delta/2$ or $G_1=G_2=G$.

%
\begin{figure}
\centering
\includegraphics[scale=0.35]{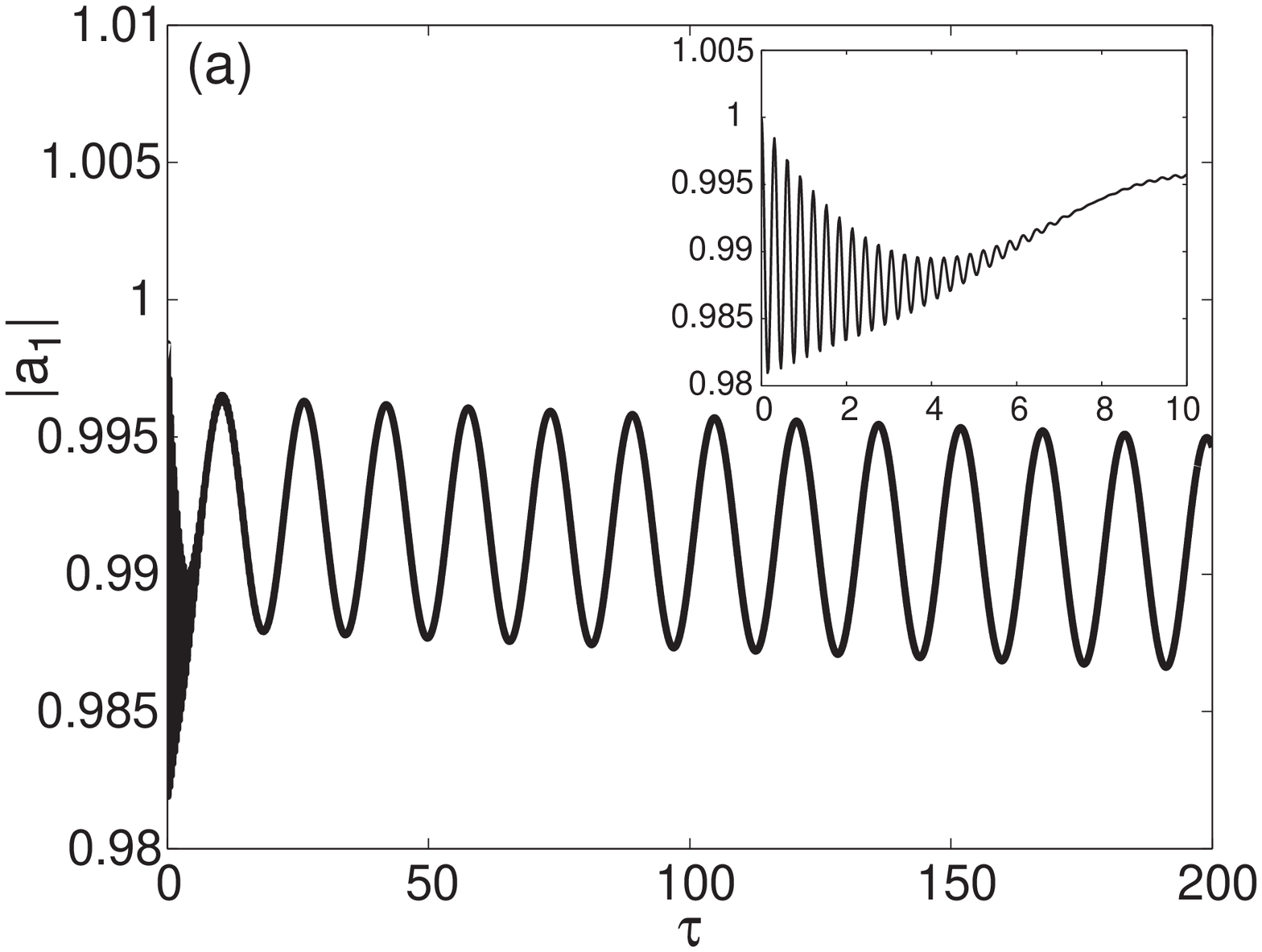}
\includegraphics[scale=0.35]{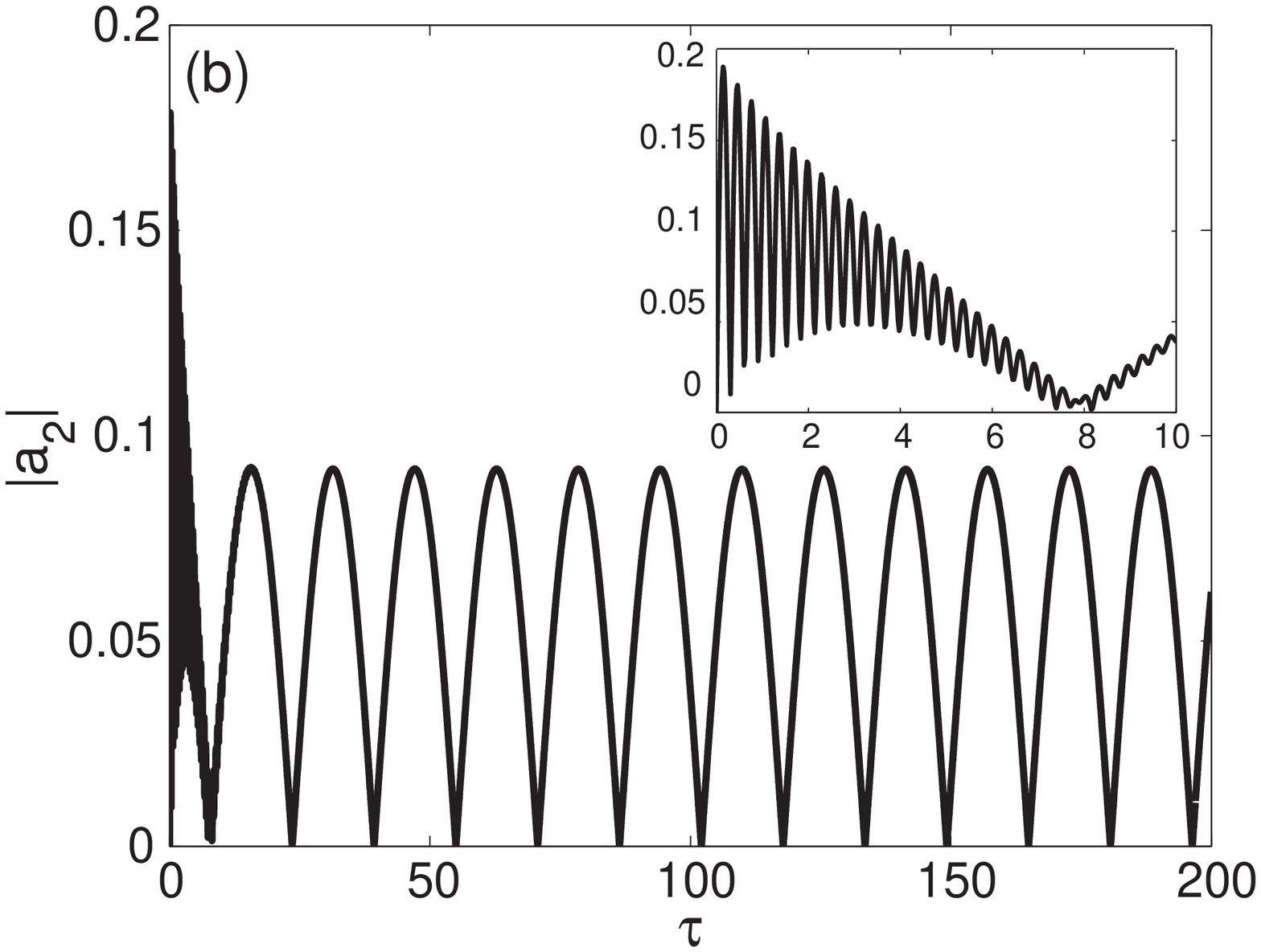}
\includegraphics[scale=0.35]{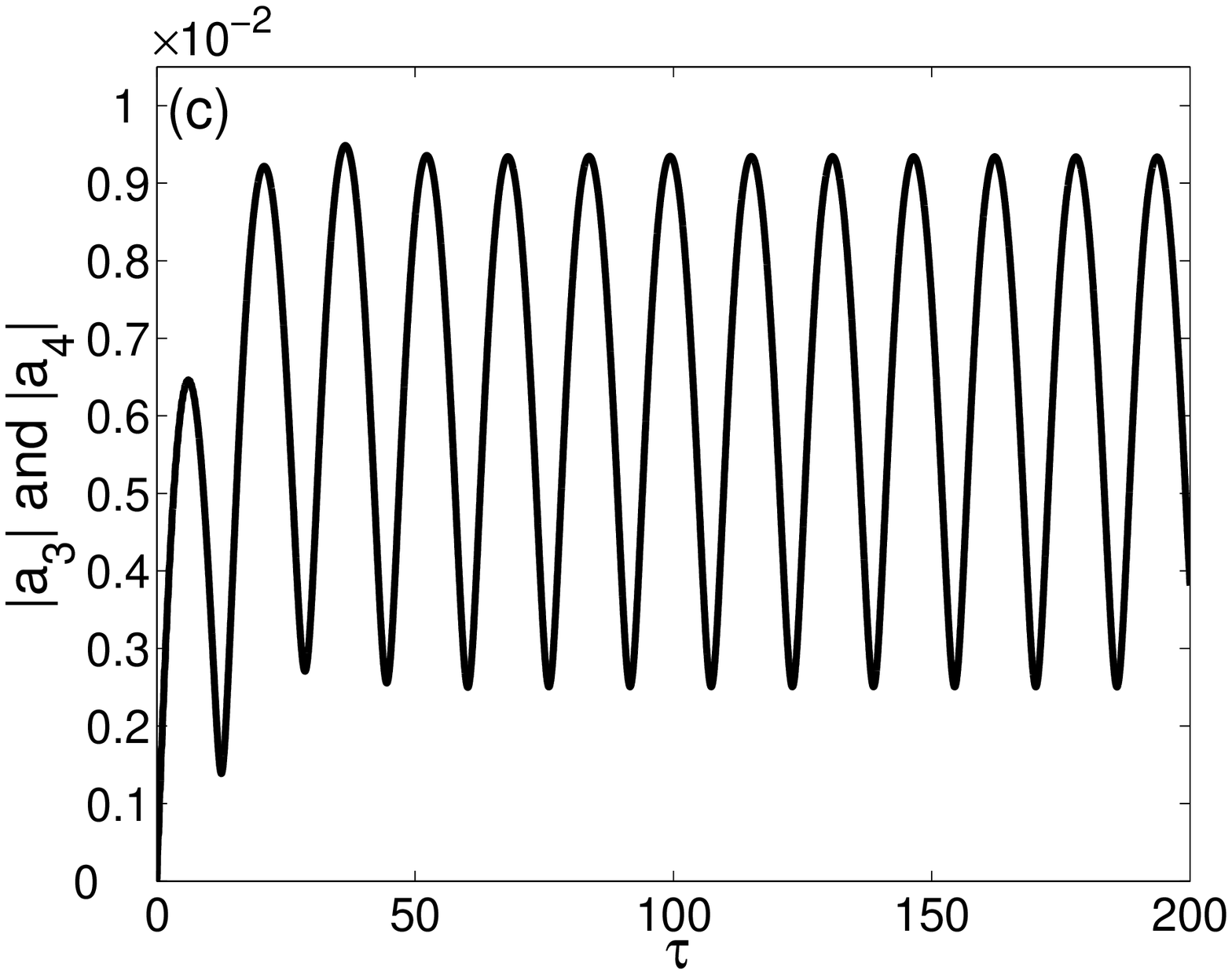}
\includegraphics[scale=0.35]{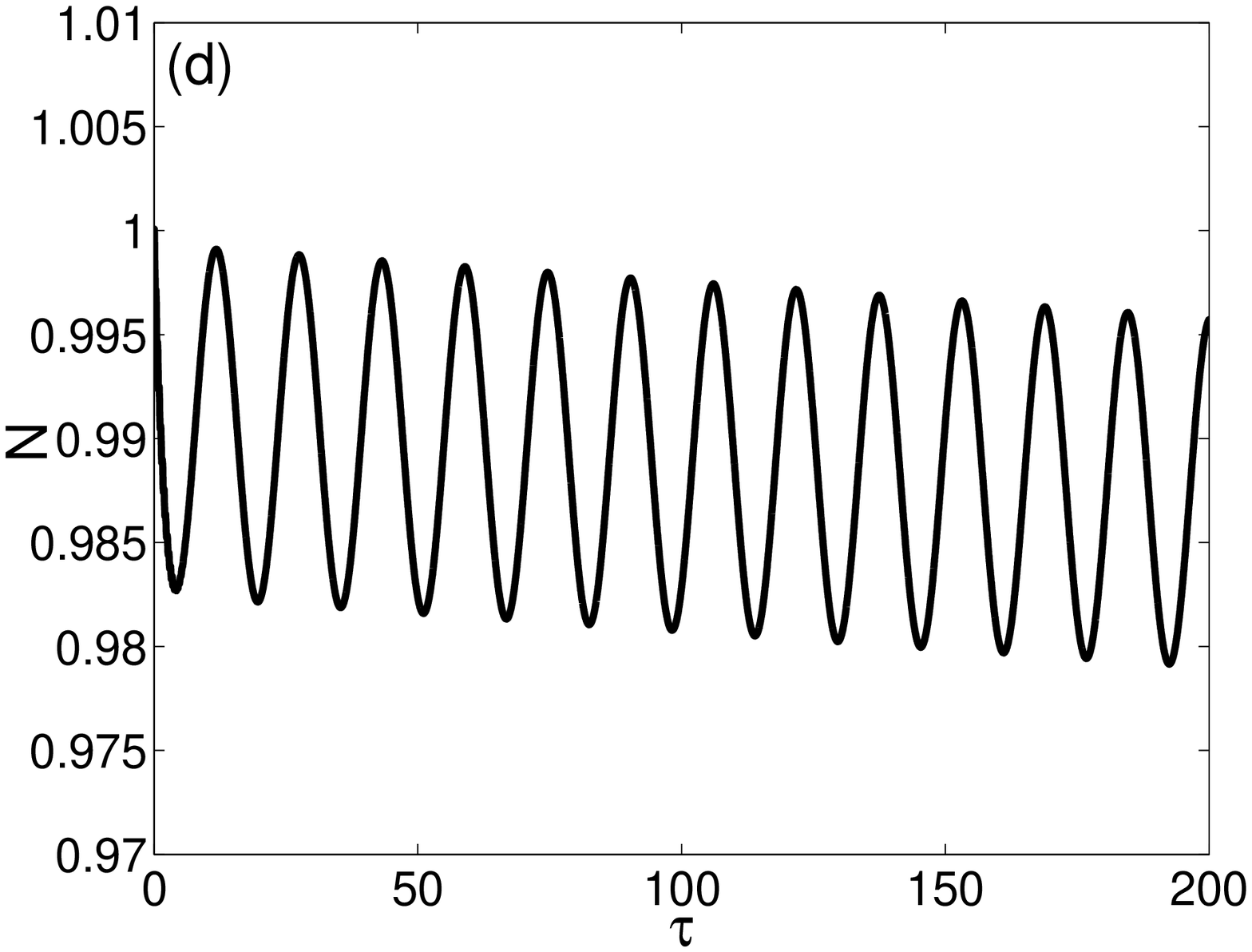}
\caption{\footnotesize The results of direct simulations of Eqs.
(\ref{AV}) with the initial conditions $a_1=1$ and
$a_2=a_3=a_4=0$. (a) The curves of $|a_1|$ versus $\tau$. The
inset shows the details for $\tau\in[0,10]$. (b) The curves of
$|a_2|$ versus $\tau$. The inset shows the details for
$\tau\in[0,10]$. (c) The curves of $|a_j|$ ($j=3,4$) versus
$\tau$. (d) The curves of $N$ versus $\tau$. Here,
$N\equiv\sum_{i=1}^{4}|a _{i}|^2$ and $\tau\equiv\Omega_{P1} t$.
The parameters are given by $\gamma_2=36$ MHz, $\gamma=10$ MHz,
$\delta_{1}=1.0\times10^{9}$ s$^{-1}$,
$\delta_{2}=1.0\times10^{7}$ s$^{-1}$, $\Delta=2.0\times10^{7}$
s$^{-1}$, $\Omega_{P1}=5.0\times10^{7}$ s$^{-1}$,
$\Omega_{P2}=5.1\times10^{7}$ s$^{-1}$, and
$\Omega_{p}=\Omega_{s}=1.0\times10^{6}$ s$^{-1}$. $\gamma_1=0.2$
MHz is obtained by Eq. (\ref{Gama}). }
\end{figure}
%
In Fig. 2, we show the results of direct simulations of Eqs.
(\ref{AV}) with a set of practical parameters given in the
caption. The initial conditions are $a_1=1$ and $a_2=a_3=a_4=0$.
The dependence of atomic probability amplitudes $a_l$ ($l$=1-4)
and quantity $\sum_{i=1}^{4}|a _{i}|^2$ on time are illustrated.
We can see that the condition $\sum_{i=1}^{4}|a _{i}|^2=1$ is
satisfied in a rather long time.

%
\begin{figure}
\centering
\includegraphics[scale=0.35]{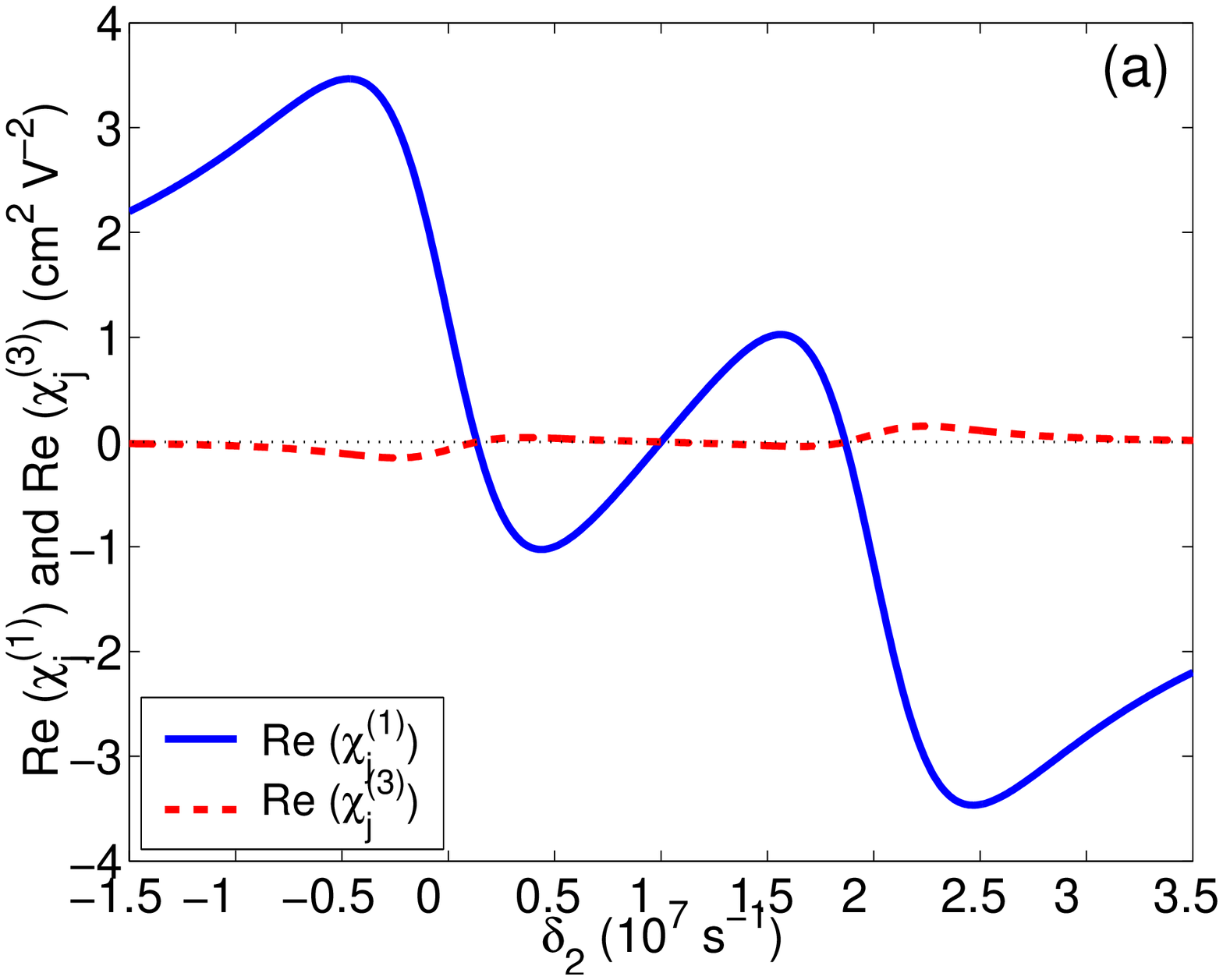}
\includegraphics[scale=0.35]{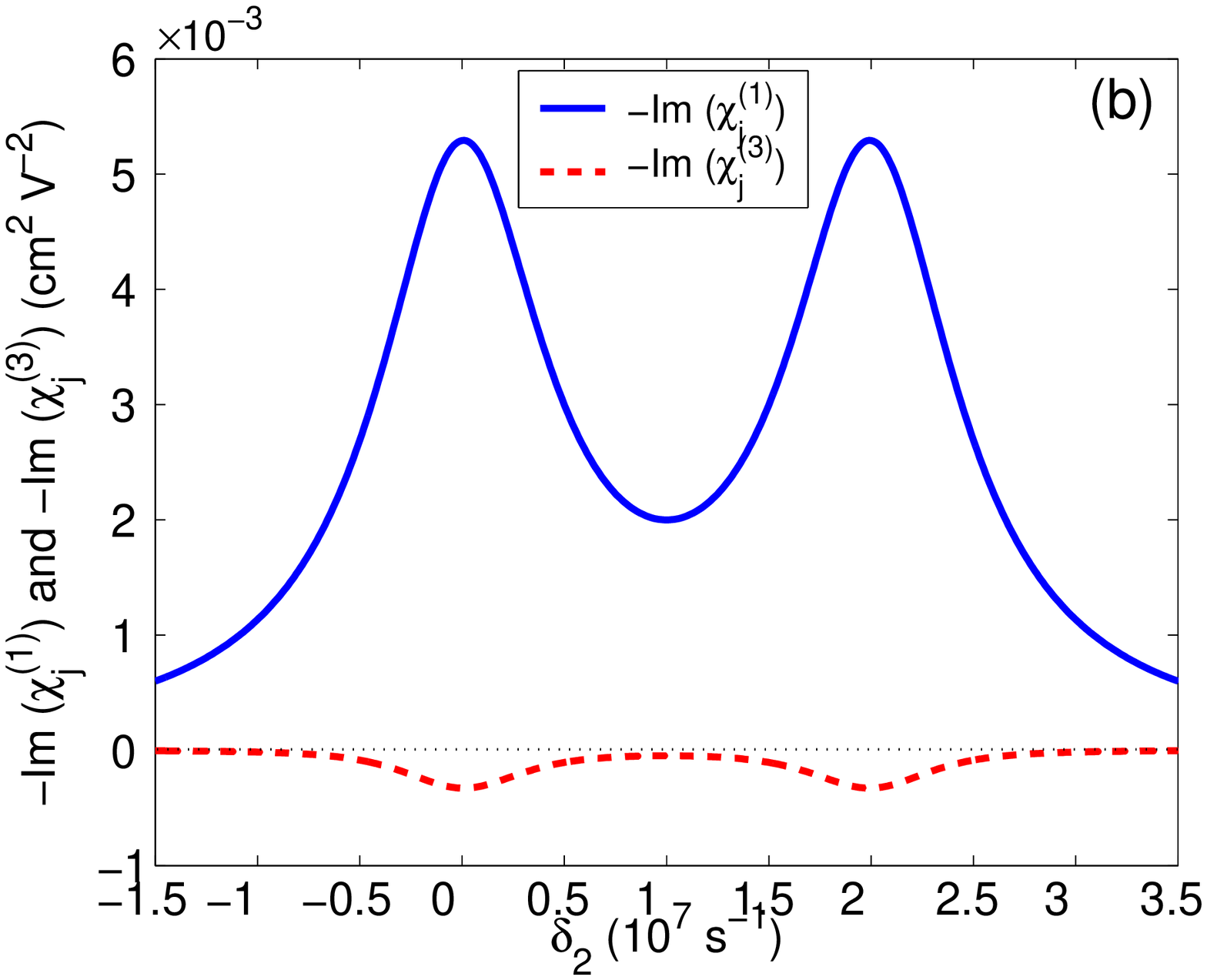}
\includegraphics[scale=0.35]{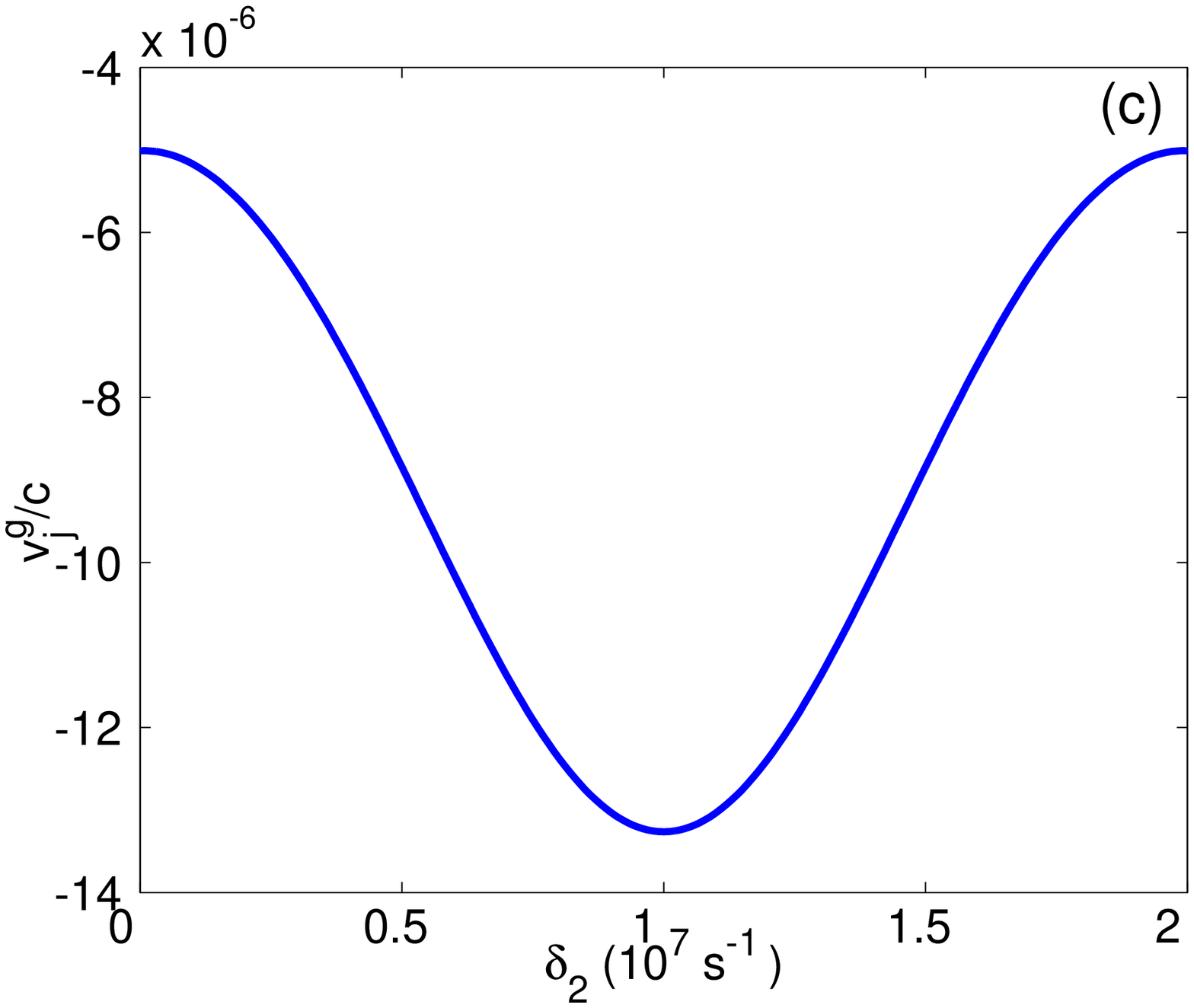}
\caption{\footnotesize (color online) (a) The curves of
Re$(\chi_j^{(1)})$ (solid line) and Re$(\chi_j^{(3)})$ (dashed line)
versus $\delta_2$. (b) The curves of  $-$Im$(\chi_j^{(1)})$ (solid
line) and $-$Im$(\chi_j^{(3)})$ (dashed line) versus $\delta_2$. (c)
The curve of $v_g^{j}/c$ versus $\delta_2$. The parameters are given
by $\gamma_2=36$ MHz, $\gamma=10$ MHz, $\delta_{1}=1.0\times10^{9}$
s$^{-1}$, $\Delta=2.0\times10^{7}$ s$^{-1}$,
$\Omega_{P1}=5.0\times10^{7}$ s$^{-1}$,
$\Omega_{P2}=5.1\times10^{7}$ s$^{-1}$, ${\cal
N}_a=1.44\times10^{13}$ cm$^{-3}$, and $D_0=2.54\times10^{-27}$ C
cm.}
\end{figure}
%
In Fig. 3(a) [Fig. 3(b)], we show the curves of Re$(\chi_j^{(1)})$
[$-{\rm Im}(\chi_j^{(1)})$] and Re$(\chi_j^{(3)})$
[$-$Im$(\chi_j^{(3)})$] versus $\delta_2$ with a set of practical
parameters given in the caption \cite{note0}. A gain doublet
structure in the spectrum can be apparently observed [see panel
(b)], where a gain minimum can be acquired at $\delta_2=\Delta/2$.
Thus, when working near the gain minimum within the hole, a rapid
increase of light intensity appeared in the ARG system can be
effectively avoided. In Fig. 3(c), we show the curves of $v_g^{j}/c$
versus $\delta_2$. The group velocity is negative (with a small
absolute value) corresponding the superluminal propagation.

Now we present the expressions of group velocities for both weak
fields, which are defined by $v_g^j=c/(1+n_g^j)$ ($j=p,s$). As we
know, the group velocities of two light pulses must be comparable in
order to achieve an effective CPM \cite{Luk}. In our system, the
group indexes of the probe and signal fields are given by
\be \label{Group}
n_{g}^{j}\simeq
-\frac{\kappa\omega_{j}}{2}\left[\frac{G_1}{\delta_2^2}
+\frac{G_2}{(\delta_2-\Delta)^2}\right].
\ee
Because in our system $\omega_p\approx \omega_s$, we have
$v_g^p\approx v_g^s$, and hence the group velocity matching is
automatically satisfied. In addition, since $n_{g}^{j}\ll-1$ (due to
the large values of $\omega_{j}$), both group velocities  are
negative, i.e., the probe and signal fields travel with superluminal
propagating velocities.

\section{Two-Qubit Polarization Phase Gates and highly entangled photons}

The prototype of optical implementation of a two-qubit gate is the
QPG in which one qubit gets a phase shift conditional to the other
qubit state according to the transformation
$|i\rangle_1|j\rangle_2\rightarrow\phi_{ij}|i\rangle_1|j\rangle_2$,
where ${i, j}=0, 1$ denote logical qubit bases. This gate becomes
universal when $\phi_{11}+\phi_{00}-\phi_{10}-\phi_{01}\neq0$
\cite{Turchette}.

We choose two orthogonal polarization states  $|\sigma^{-}\rangle$
and $|\sigma^{+}\rangle$ to encode binary information for each
qubit. The scheme shown in Fig. 1 is completely implemented only
if both probe and signal fields have the ``right'' polarization
states. When both of two weak fields have ``wrong'' polarization
states, there is no sufficiently close excited state to which
levels $|3\rangle$ and $|4\rangle$ can couple, and hence the probe
and signal fields will only acquire the trivial vacuum phase shift
$\phi_{0}^j=k_jL$. Here $k_j\equiv \omega_j/c$ ($j=p,s$), and  $L$
denotes the length of the medium. When one of the two weak fields
have ``wrong'' polarization state, say for a
$\sigma^{-}$-polarized probe field, there is no sufficiently close
excited state to which levels $|3\rangle$ can couple and the
signal field subjects to the $\Lambda$-configuration constituted
by $|1\rangle$, $|2\rangle$, and $|4\rangle$ levels. Thus the
signal field experiences a self-Kerr effect and acquires a
nontrivial phase shift $\phi_1^s$, while the probe field acquires
only a vacuum phase shift $\phi_0^{p}$. When only the probe and
the signal fields have ``right'' polarization states, they all
acquire nontrivial phase shifts  $\phi_2^p$ and $\phi_2^s$,
respectively.

Assume that the input probe and signal pulses can be treated as
polarized single photon wave packets, expressed as a superposition
of the circularly polarized states, i.e.
$|\psi\rangle_j=1/\sqrt{2}|\sigma^-\rangle_j+1/\sqrt{2}
|\sigma^+\rangle_j$ ($j=p,s$). Here $|\sigma^\pm\rangle_j=\int
d\omega \xi_j(\omega) a_\pm^\dagger(\omega)|0\rangle$ with
$\xi_j(\omega)$ being a Gaussian frequency distribution of
incident wave packet centered at frequency $\omega_j$. The photon
field operators undergo a transformation while propagating through
the atomic medium of length $L$, i.e. $a_\pm(\omega)\rightarrow
a_\pm(\omega)\exp\{i\omega/c\int_{0}^{L}dz n_\pm(\omega,z)\}$.
Assuming $n_\pm(\omega,z)$ (the real part of the refractive index)
varies slowly over the bandwidth of the wave packet centered at
$\omega_j$, one gets
$|\sigma^\pm\rangle_j\rightarrow\exp{(-i\phi_\pm^j)}|\sigma^\pm\rangle_j$,
with $\phi_\pm^j=\omega_jn_\pm(\omega_j,z)L/c$. Thus, the truth
table for a polarization two-qubit QPG using the present
configuration is given by
\bes \label{truth}
\bea
& & \label{truth1}
|\sigma^-\rangle_p|\sigma^+\rangle_s\rightarrow\exp{[-i(\phi_0^p+\phi_0^s)]}
|\sigma^-\rangle_p|\sigma^+\rangle_s,\\
& & \label{truth2}
|\sigma^-\rangle_p|\sigma^-\rangle_s\rightarrow\exp{[-i(\phi_0^p+\phi_1^s)]}
|\sigma^-\rangle_p|\sigma^-\rangle_s,\\
& & \label{truth3}
|\sigma^+\rangle_p|\sigma^+\rangle_s\rightarrow\exp{[-i(\phi_1^p+\phi_0^s)]}
|\sigma^+\rangle_p|\sigma^+\rangle_s,\\
& & \label{truth4}
|\sigma^+\rangle_p|\sigma^-\rangle_s\rightarrow\exp{[-i(\phi_2^p+\phi_2^s)]}
|\sigma^+\rangle_p|\sigma^-\rangle_s,
\eea
\ees
where $\phi_0^{j}=k_{j}L$,
$\phi_1^{j}=k_{j}L(1+2\pi\chi_{j}^{(1)})+\phi^{(j,s)}$, and
$\phi_2^{j}=\phi_1^{j}+\phi^{(j,c)}$, with
\bes \label{phi3}
\bea
& & \label{phi31} \phi^{(j,s)}=k_{j}L\frac{\pi^{3/2}\hbar^2|
\Omega_{j}|^2}{4|D_{2}|^2}{\rm Re}[\chi_{j}^{(3,s)}],\\
& & \label{phi32}\phi^{(j,c)}=k_{j}L\frac{\pi^{3/2}\hbar^2|
\Omega_{j'}|^2}{4|D_{2}|^2}{\rm Re}[\chi_{j}^{(3,c)}]
\frac{\text{erf}(\xi_{jj'})}{\xi_{jj'}},
\eea
\ees
contributed respectively by self-phase modulation (SPM) and
cross-phase modulation (CPM), where
$\xi_{jj'}=\sqrt{2}L(1-v_g^j/v_g^{j'})/(\tau_{j'}v_g^j)$, with
$\tau_{j'}$ being the width of the pulse. If group velocity
matching is satisfied, i.e. $\xi_{jj'}\rightarrow 0$,
$\text{erf}[\xi_{jj'}]/\xi_{jj'}$ reaches its maximum value
$2/\sqrt{\pi}$.

From Eq. (\ref{truth}), we can compute the degree of entanglement
of the two-qubit state by using the entanglement of formation. For
an arbitrary two-qubit system, it is given by \cite{Wootters}
\be E_F(C)=h\left( \frac{1+\sqrt{1-C^2}}{2}\right), \ee
where $h(x)=-x{\rm log}_2(x)-(1-x){\rm log}_2(1-x)$ is Shannon's
entropy function, $C$ is the concurrence given by
$C(\hat{\rho})={\rm max}\{0,
\lambda_1-\lambda_2-\lambda_3-\lambda_4\}$. Here $\lambda_i$'s are
square roots of eigenvalues of the matrix
\be \label{dens}
\hat{\rho}\tilde{\hat{\rho}}=\hat{\rho}
\hat{\sigma}_y^p\otimes\hat{\sigma}_y^s\hat{\rho}^*
\hat{\sigma}_y^p\otimes\hat{\sigma}_y^s
\ee
in decreasing order. The density matrix $\hat{\rho}$ in Eq.
(\ref{dens}) can be directly obtained by using Eq. (\ref{truth}),
the quantity $\tilde{\hat{\rho}}$ ($\hat{\rho}^*$) means the
transpose (complex conjugation) of $\hat{\rho}$, and
$\hat{\sigma}_y$ denotes the $y$-component of the Pauli matrix.

Eq. (\ref{truth}) supports a universal QPG if the conditional
phase shift \cite{Turchette}
\bea \label{Univ}
& & (\phi_0^p+\phi_0^s)+(\phi_2^p+\phi_2^s)-(\phi_0^p+\phi_1^s)-(\phi_1^p+\phi_0^s)\nonumber\\
& & =\phi^{(p,c)}+\phi^{(s,c)}
\eea
is non-zero. From this formula, we see that only the phase shifts
due to the CPM effect contribute to the conditional phase shifts.

Now we provide a practical set of parameters corresponding to
typical values of $^{87}$Rb atoms in room temperature. The decay
rate of the lower states, i.e. $|3\rangle$ ($5^{2}S_{1/2}$,
$F=2,m=-1$) and $|4\rangle$ ($5^{2}S_{1/2}$, $F=2,m=1$), is
$\gamma=300$ Hz. The hyperfine splitting between the lower states
can be adjusted by the intensity of an externally applied magnetic
field. For a magnetic field $\approx340$ G we obtain the splitting
$\approx3.8$ GHz. The decay rate of the higher state $|2\rangle$
($5^{2}P_{1/2}$, $F=2,m=0$) is $\gamma_2=36$ MHz. The other
parameters are taken the same with those used in Fig. 2, as well as
$\delta_2=0.8\times10^{7}$ s$^{-1}$. With the given parameters, we
obtain that $\chi_j^{(1)}=-0.10\times10^{-2}-i0.85\times10^{-7}$ and
$\chi_j^{(3)}=0.34\times10^{-4}+i0.28\times10^{-8}$ cm$^{2}$
V$^{-2}$. We note that the imaginary parts of the susceptibilities
are much smaller than those of the real parts due to the conditions
$\gamma_2\ll\delta_1$, $\gamma\ll\delta_2$, and
$\delta_2\neq\Delta/2$. A very small total gain effect remains after
the balance of the linear gain and nonlinear absorption. The real
parts of the third-order susceptibilities is about $\sim10^{13}$
times larger than those measured for usual nonlinear optical
materials, i.e., a giant enhancement of CPM can be achieved in our
system. The group velocities of the both probe and signal fields are
very well matched, with the values
\be
v_g^p\approx v_g^s = -0.94\times10^{-5}c,
\ee
corresponding to indeed a superluminal propagation.

In Fig. 4(a), we have shown the calculating result of degree of
entanglement versus the propagation the device length $L$. We see
that a nearly $100\%$ degree of entanglement can be obtained at
$L=0.53$ cm. The reason for acquiring such a high degree of
entanglement is due to the non-absorption feature of the system.
Shown in Fig. 4(b) are the curves of CPM induced phase shifts
$\phi^{(p,c)}$ and $\phi^{(s,c)}$ versus $L$. We see that a
conditional phase shift $\phi^{(p,c)}+\phi^{(s,c)}$ up to $\pi$
radians can be obtained at $L\simeq0.53$ cm, corresponding to the
point of the maximum entanglement in Fig. 4(a). In Fig. 4(c), we
show the curves of $\phi^{(p,c)}$ and $\phi^{(s,c)}$ versus
$\delta_2$ at $L=0.53$ cm.

The probe and signal fields can have a mean amplitude of about one
photon when these beams are focused or propagate in a tightly
confined waveguide (e.g. hollow-core photonic crystal fibers
\cite{baj}). With the above parameters, we obtain the intensities
of the probe ($I_p$) and signal ($I_s$) fields,  given by
$I_p\approx I_s= 0.23\times 10^{-6}$ W cm$^{-2}$ when
$\Omega_p\approx \Omega_s= 1.0\times10^6$ s$^{-1}$. We remark that
the intensity of a single 800-nm photon per nanosecond on the area
of 1 $\mu$m$^2$ is $I_{ph}=2.5\times10^{-2}$ W cm$^{-2}$. This
shows that our scheme can indeed make a polarization QPG with
$\pi$ conditional phase shift possible with single photon wave
packets. Based on the superluminal propagating velocities and the
enhanced CPM, the probe and signal fields acquire nontrivial
nonlinear phase shifts when both of them have ``right''
polarization states in a fast response time and a short
propagation distance, which allow us to implement a rapidly
responding phase gate. For instance, if the group velocity of the
probe and signal waves are reduced $10^{-4} c$ when using the
EIT-based scheme, these waves will take around 180 ns  to pass the
device (for $L=0.53$ cm) during which the nonlinear phase-shifting
probe and signal fields must be present all the time. However, to
acquire the same amount of the nonlinear phase shift for the probe
and signal waves in the present ARG system, the device transient
time is only about 18 ps \cite{note1}.

\begin{figure}
\includegraphics[scale=0.35]{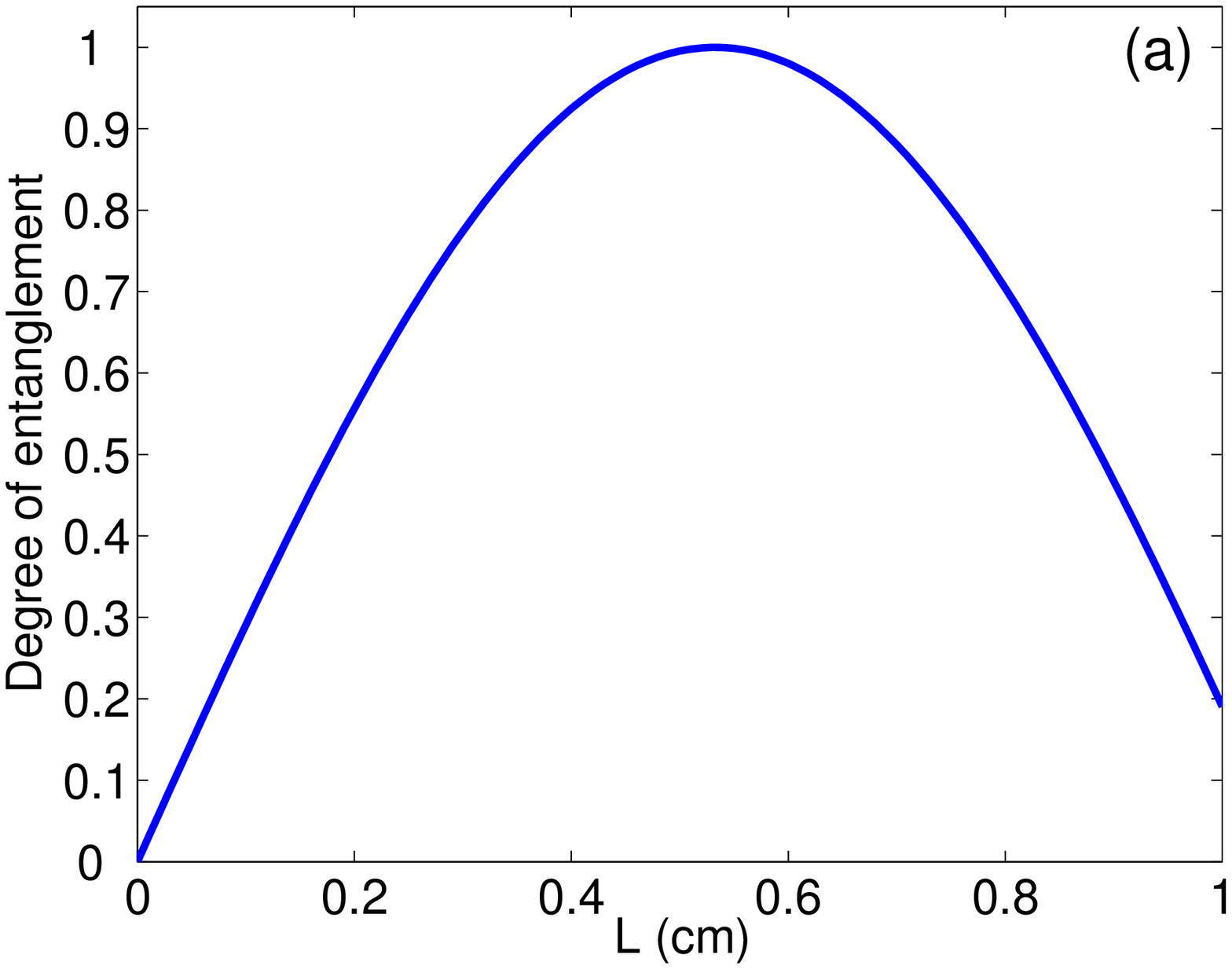}
\includegraphics[scale=0.35]{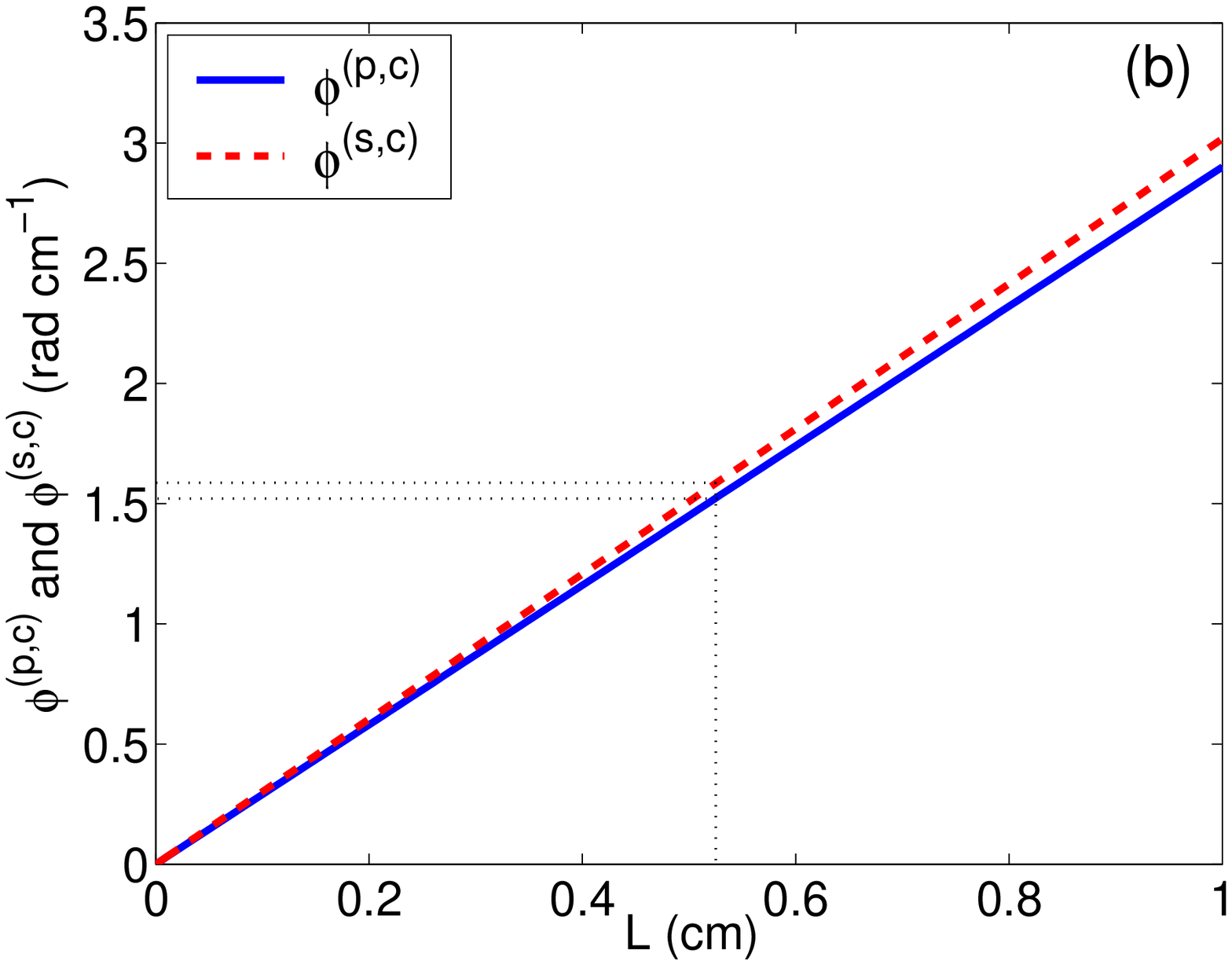}
\includegraphics[scale=0.35]{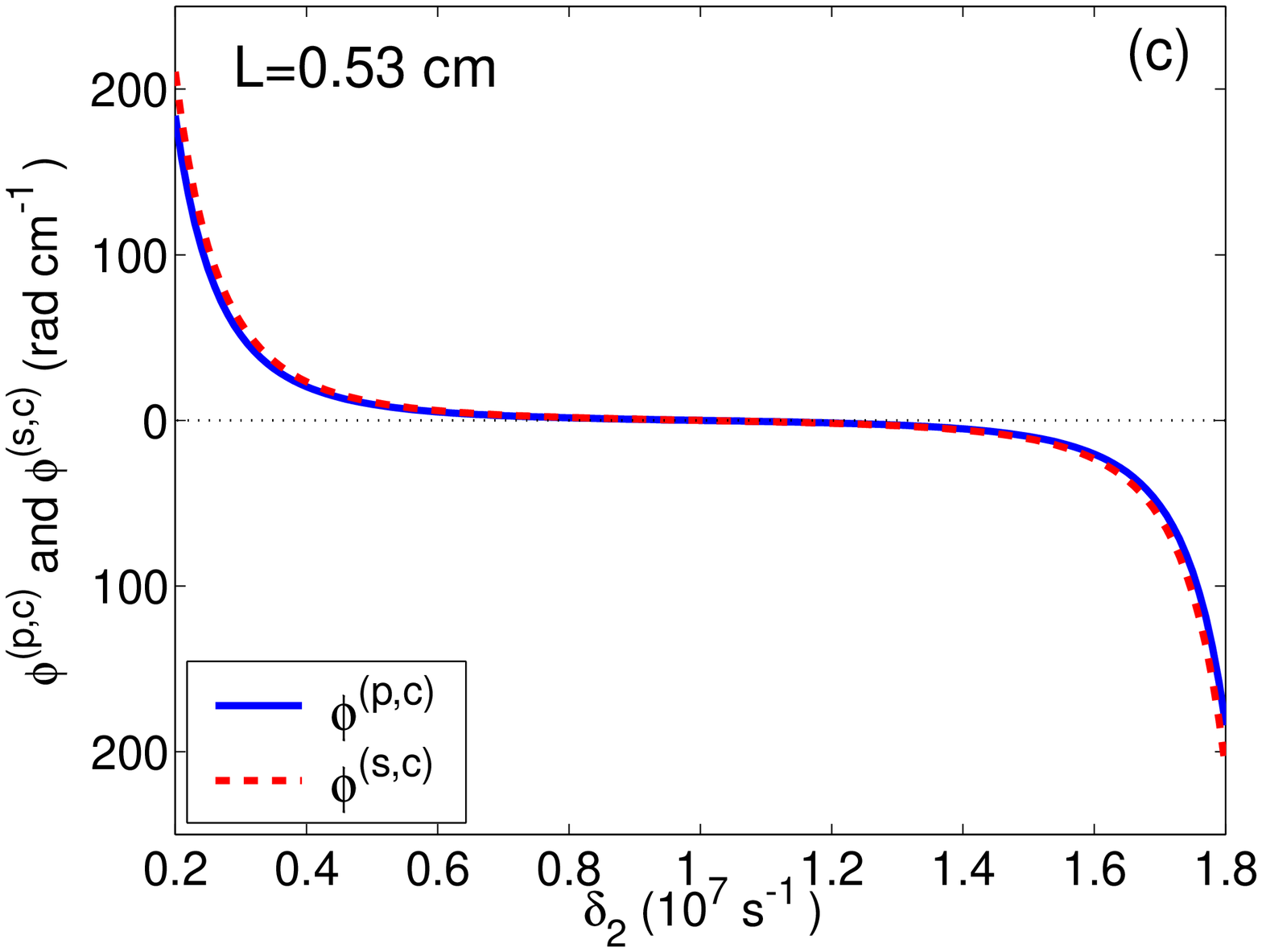}
\centering \caption{\footnotesize (color online) (a) The degree of
entanglement versus the device length $L$. (b) The curves of
$\phi^{(p,c)}$ and $\phi^{(s,c)}$ versus $L$. A conditional phase
shift $\phi^{(p,c)}+\phi^{(s,c)}$ up to $\pi$ radians can be
obtained at $L\simeq0.53$ cm. (c) The curves of $\phi^{(p,c)}$ and
$\phi^{(s,c)}$ versus $\delta_2$ at $L=0.53$ cm. The parameters
are given in the text. }
\end{figure}

\section{Discussion and summary}

Now we briefly discuss the Doppler effect due to the atom's
thermal motion. Actually, our results can be readily generalized
when an atom moves with a velocity $V$ by the replacement
$\delta_1\rightarrow\omega_{21}-\omega_{P1}+k_{P1z}V_z$,
$\delta_2\rightarrow(\omega_{P1}-\omega_p)-\omega_{31}+(k_p-k_{P1})_zV_z
=(\omega_{P1}-\omega_s)-\omega_{41}+(k_s-k_{P1})_zV_z$ (we assume
all light fields propagate along the $z$-direction, as suggested
in the inset of Fig. 1). The $V_z$-dependent terms obtained are
then averaged over a given thermal velocity distribution $f(V_z)$.
From the above discussions, we see that the velocity-dependent
effect in the two-photon detunings $\delta_2$ in the copropagating
case $k_{P1}k_j>0$ is much smaller than those in the
counterpropagating case $k_{P1}k_j<0$. Consequently, the
velocity-dependent effect or the Doppler effect in the two-photon
detunings can be usually neglected compared with that in the
one-photon detuning if we choose the waves to propagate in the
same direction. Moreover, such effect in the one-photon detuning
can also be efficiently suppressed if
$\omega_{21}-\omega_{P1}\gg\omega_{P1z}V_z$, which is satisfied in
our system.

The experimental demonstration of the phase gate requires the
measurement of phase shifts, which will result in errors due to the
fluctuations of light intensities and frequency detunings of the
probe and signal fields. In order to minimize the effect of relative
detuning fluctuations, one can take all lasers tightly phase locked
to each other. The light intensity having fluctuations of $1\%$ will
yield an error less than $2\%$ in the phase measurement.

We should point out that although CPM is a very promising
candidate for the design of deterministic optical quantum phase
gates, it still faces some challenges, which include: (i) How to
achieve the sufficiently high single-photon intensity; (ii) How to
overcome the phase noise induced by non-instantaneous nonlinear
response inherent in resonant atomic systems; (iii) How to obtain
a spatially homogeneous CPM necessary for effective entanglement
between light pulses, etc. These problems are now actively
investigated, and some methods for dealing with them have already
been proposed \cite{mar}. On the other hand, in the present work
we have treated the probe and signal fields in a classical way.
Therefore, one would be easier to create the entanglement of
macroscopic, coherent states rather than single photon states. A
full quantum treatment is still necessary but beyond the scope of
the present work.

To sum up, we have presented a scheme for obtaining  entangled
photons and quantum phase gates in a room-temperature four-state
tripod-type atomic system with a two-mode ARG. We have analyzed the
linear and nonlinear optical response of the ARG system and shown
that the scheme is fundamentally different from those based on
electromagnetically induced transparency and hence can avoid
significant probe-field absorption as well as temperature-related
Doppler effect. We have demonstrated that highly entangled photon
pairs can be produced and rapidly responding polarization qubit
quantum phase gates can be constructed based on the unique features
of enhanced cross-phase modulation and superluminal probe-field
propagation of the ARG system. The method provided here can also be
extend to the study on multi-way entanglement and multi-qubit phase
gates.

\section*{ACKNOWLEDGMENTS}

Authors thank Y. Li and L. Deng for useful discussions and
suggestions. This work was supported by NSF-China under Grants No.
10434060 and No. 10874043, by the Key Development Program for
Basic Research of China under Grants No. 2005CB724508 and No.
2006CB921104. C. H. was supported by the Funda\c{c}\~ao para a
Ci\^encia e a Tecnologia (FCT) under Grant No. SFRH/BPD/36385/2007
and the Calouste Gulbenkian Foundation 2009.




\begin{references}
\bibitem{note} If the group velocity of a light pulse is reduced by a factor of $F$
from the speed of light in vacuum, the phase shift operation will
be a factor of $F$ slower \cite{Deng}.
\bibitem{note0}Due to the symmetric configuration of the system,
the linear and nonlinear susceptibilities of both the probe and
signal fields are almost equal (see Eq. (\ref{Chi2})\,). Hence, the
curves plotted in Fig. 3 are valid for both the probe and signal
fields.
\bibitem{note1}The lead time is typically about $10\%$ of $c$.
Thus the device transient time can be estimated quite accurately
by using free-space speed of light \cite{jia1}.
\end{references}
\end{document}